\documentclass[aps,prb,twocolumn,superscriptaddress,amsfonts,amssymb,amsmath,floatfix]{revtex4}

\usepackage[T1]{fontenc}
\usepackage[utf8]{inputenc}
\usepackage{amsmath} 
\usepackage{amsfonts}
\usepackage{amssymb}
\usepackage{graphicx} 
\usepackage[english]{babel} 
\usepackage{hyperref}

\usepackage{todonotes}

\begin{document}

\title{Control of a polar order via magnetic field in a vector-chiral magnet}

\author{Martina Dragi\v{c}evi\'{c}}
\email{mdragicevic@ifs.hr}
\author{David \surname{Rivas G\'{o}ngora}}
\author{\v{Z}eljko Rapljenovi\'{c}}
\author{Mirta Herak}
\author{Vedran Brusar}
\author{Damir Altus}
\affiliation{Institute of Physics, Bijeni\v{c}ka cesta 46, HR-10000 Zagreb, Croatia}
\author{Matej Pregelj}
\author{Andrej Zorko}
\affiliation{Jo\v{z}ef Stefan Institute, Jamova c.\ 39, 1000 Ljubljana, Slovenia}
\author{Helmuth Berger}
\affiliation{Ecole polytechnique f\'{e}d\'{e}rale de Lausanne, CH-1015 Lausanne, Switzerland}
\author{Denis Ar\v{c}on}
\affiliation{Jo\v{z}ef Stefan Institute, Jamova c.\ 39, 1000 Ljubljana, Slovenia}
\affiliation{Faculty of Mathematics and Physics, University of Ljubljana, Jadranska c.\ 19, 1000 Ljubljana, Slovenia}
\author{Tomislav Ivek}
\affiliation{Institute of Physics, Bijeni\v{c}ka cesta 46, HR-10000 Zagreb, Croatia}


\date{\today}

\begin{abstract}
Vector-chiral (VC) antiferromagnetism is a spiral-like ordering of spins which may allow ferroelectricity to occur due to loss of space inversion symmetry. In this paper we report direct experimental observation of ferroelectricity in the VC phase of $\beta$-TeVO$_4$, a frustrated spin chain system with pronounced magnetic anisotropy and a rich phase diagram. Saturation polarization is proportional to neutron scattering intensities that correspond to the VC magnetic reflection. This implies that inverse Dzyaloshinskii-Moriya mechanism is responsible for driving electric polarization. Linear magnetoelectric coupling is absent, however an unprecedented dependence of electric coercive field on applied magnetic field reveals a novel way of manipulating multiferroic information.
\end{abstract}


\maketitle 


Multiferroics are materials with simultaneous magnetic and ferroelectric order, often with some weak coupling between the two orderings with a different type of symmetry. It has been shown that spin order may initiate ferroelectric order.\cite{Kimura2003} Still, magnets with a confirmed ordering of electric dipoles remain scarce and mechanisms of their formation are under active research.\cite{Hill2000, Eerenstein2006, Spaldin2019} Evidence has been mounting that certain antiferromagnetic phases with spiral arrangements of spins may break space inversion symmetry and allow electric dipoles to form.\cite{ruff2019chirality, mochizuki2015dynamical, Cheong2007, Mostovoy2006, Arima2011, Krohns2019, Yamasaki2007, Terada2008} However, controlling ferroelectric properties of these phases with magnetic field remains a challenge.

The rich phase diagram of $\beta$-TeVO$_4$ comprises several magnetically ordered phases and may provide a platform to explore emergent ferroelectric properties. This system is an insulating quasi-one-dimensional magnet with two zig-zag spin-chains along the crystallographic $c$-axis and belongs to the $P2_1/c$ space group.\cite{meunier73} Notably, vanadium spins (V$^{4+}$, $S = 1/2$) interact via nearest ferromagnetic (V-O-V) and next-nearest antiferromagnetic (V-O-Te-O-V) superexchange.\cite{savina2011,Saul2014,pregelj2015, Savina2015b,weickert2016} The resulting frustration is amplified by quantum fluctuations giving rise to three distinct magnetic phase transitions at low temperatures.\cite{savina2015} In the absence of external magnetic field, at $T_\textrm{N1}=4.65$\,K the paramagnetic phase gives way to an incommensurate spin-density wave (SDW).\cite{savina2015,pregelj2015,weickert2016} Then, at $T_\textrm{N2}=3.28$\,K a second incommensurate, orthogonal SDW appears with a different periodicity and the two modulations form a distinct spin-stripe phase which is unique among multiferroics.\cite{pregelj2015,pregelj2016,pregelj2019elementary} The difference between the two magnetic wave vectors is imposed by magnetic exchange anisotropy, which is responsible also for their different temperature dependencies. As temperature is lowered, this difference is reduced and at $T_\textrm{N3}=2.28$\,K the two wave vectors become equal and locked. There, the vector-chiral (VC) ground state is established and on each of the two vanadium chains a spin spiral appears, one with spins in the $ab$ plane and the other in the $bc$ plane.\cite{pregelj2016,weickert2016,pregelj2019magnetic} Application of low external magnetic fields gives rise to a strongly anisotropic phase diagram.\cite{savina2015, weickert2016, pregelj2019magnetic, herak2020} Magnetic field $\mathbf{H}||b$ stabilizes the VC phase, suppresses the spin-stripe phase and increases $T_\mathrm{N3}$. On the other hand, $\mathbf{H}{\perp}b$ suppresses the VC phase which disappears above about 2\,T.

Recently, a peak in dielectric response was found at $T_\mathrm{N3}$ where $\beta$-TeVO$_4$ enters the VC phase. This suggest the formation of electric dipoles and perhaps their ferroelectric ordering.\cite{pregelj2019elementary} The possibility of electric dipoles is additionally supported by indications of structural deformation occuring at $T_\mathrm{N3}$.\cite{weickert2016}

In this work, we present firm experimental evidence of ferroelectric ordering within the vector-chiral phase of single-crystal $\beta$-TeVO$_4$. By measuring static electric polarization and ac dielectric response, we detect the onset of ferroelectricity exactly at the phase transition into the VC magnetic phase. The appearance of VC ordering breaks the spatial inversion symmetry and we propose that an inverse-Dzyaloshinskii-Moriya-type interaction generates electric dipoles oriented along the $b$-axis, i.e., perpendicular to the chains. We find that the saturation polarization of $\beta$-TeVO$_4$ follows the neutron scattering intensity of the VC magnetic reflection, i.e., the magnetic order parameter. The chiral order in this phase does not change with the applied magnetic field and shows no linear magnetoelectric effect, yet we observe a remarkably strong dependence of the electric coercive field on the applied magnetic field. This points to magnetic-field-enhanced pinning of the VC-ferroelectric domains, which must be caused by a magnetoelectric term of a higher order. Our results suggest that magnetic manipulation of the chiral energy landscape may have potential uses for control and stabilization of dense ferroelectric information.

\begin{figure*}
    \includegraphics[width=1.0\linewidth]{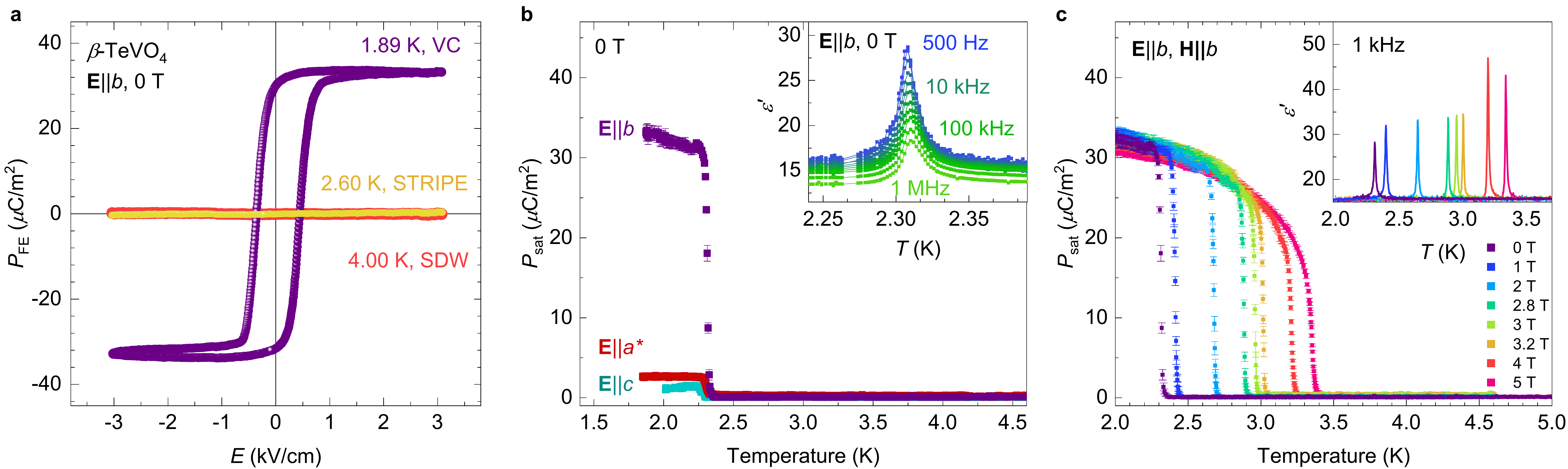}
    \caption{Static electric polarization and dielectric properties of single-crystal $\beta$-TeVO$_4$ measured using modified Sawyer-Tower-type circuit\cite{Prume2004} and LCR meter, respectively. (a) Ferroelectric contribution to the static electric polarization for $\mathbf{E}||b$ as a function of electric field $E$ at 0\,T within the SDW, spin-stripe, and VC phase. (b) Anisotropy of ferroelectric saturation polarization in 0\,T as a function of temperature. Inset shows the temperature dependence of ac dielectric response for $\mathbf{E}||b$ in linear regime for various frequencies. (c) Saturation polarization for $\mathbf{E}||b$ and dielectric response (inset) as a function of temperature at various magnetic fields $\mathbf{H}||b$.}
    \label{fig:diel}
\end{figure*}

We focus on static electric polarization and ac dielectric response of single crystal $\beta$-TeVO$_4$ in applied magnetic field. High-quality single crystals of $\beta$-TeVO$_4$ were synthesized via chemical vapor transport reaction and were cut to plate-like samples with edges along the crystal axes.\cite{pregelj2015,meunier73} Static electric polarization hysteresis $P$ \textit{vs} $E$ ($\mathbf{E}||a^\ast$, $b$, and $c$) and dielectric response of $\beta$-TeVO$_4$ were measured using modified Sawyer-Tower-type circuit\cite{Prume2004} and Keysight E4980AL-102 LCR meter, respectively, at temperatures down to 1.8\,K and in magnetic fields up to 5\,T. See Supplemental Material for details on sample preparation, electric polarization, dielectric response, and magnetic torque measurements with applied electric field.

Fig.\ \ref{fig:diel}(a) shows three selected polarization loops (dielectric linear contribution was subtracted) for $\mathbf{E}||b$ direction at 0\,T within the SDW phase, spin-stripe phase, and the VC phase. Above $T_\mathrm{N3}$, no ferroelectric-like contribution is visible. In contrast, below $T_\mathrm{N3}$ a well-defined ferroelectric hysteresis loop appears with the saturation field of about 1\,kV/cm and the saturation polarization $P_\textrm{sat} \simeq 30$\,$\mu$C/m$^2$, comparable to other spin-spiral-induced ferroelectrics (see Ref. \onlinecite{wang2009}, Table II). 

Fig.\ \ref{fig:diel}(b) shows the temperature dependence of anisotropy of saturation polarization $P_\textrm{sat}$ at 0\,T. Spontaneous electric polarization strongly prefers $\mathbf{P}||b$ as it is at least one order of magnitude larger than for $a^\ast$ and $c$ direction, the latter being the direction of chains. This is reminiscent of multiferroic TbMnO$_3$ where ferroelectric polarization perpendicular to spin chains appears with the spiral order.\cite{Kenzelmann2005} Inset in Fig.\ \ref{fig:diel}(b) shows the real part of the low-frequency dielectric function, $\varepsilon'$, for $\mathbf{E}||b$ at 0\,T. Away from $T_\mathrm{N3}$, $\varepsilon'\sim 16$ does not depend on temperature. At the transition into the VC phase, a pronounced Curie-Weiss-like peak appears and reaches maximum value of $\varepsilon'_\mathrm{max} \simeq 18$--30. Such a peak indicates ferroelectric ordering of static electric dipoles in the mean-field fluctuating local electric fields.\cite{Schulze1963} A significantly weaker peak for $\mathbf{E}||a^\ast$ at 0\,T, was reported in our previous work.\cite{pregelj2019elementary} The frequency-independent temperature position of the peak rules out relaxor-like or glassy processes and is similar to TbMnO$_3$ close to the Curie point.\cite{Trepakov2016_2} It is interesting that, in contrast to $\beta$-TeVO$_4$, the chiral spin order of the structurally related h\"{u}bnerite MnWO$_4$ shows a pronounced dielectric dispersion in frequency which indicates slow glass-like relaxation processes.\cite{Niermann2014,Niermann2015} We find the dielectric response of $\beta$-TeVO$_4$ to be without dissipation, $\varepsilon''=0$, except in a narrow temperature window of $\pm20$\,mK in vicinity of $T_\mathrm{N3}$ (See Suplementary Material).

Fig.\ \ref{fig:diel}(c) shows that the appearance of the ferroelectric polarization is pushed to higher temperatures as the VC phase is promoted with applied $\mathbf{H}||b$. We find a similar anisotropy of polarization as for 0\,T (see Supplemental Material). Deeper within the VC phase, saturation polarization does not change significantly with magnetic field. The accompanying dielectric peak also follows the same temperature dependence [inset of Fig.\ \ref{fig:diel}(c)]. In particular, it remains approximately unchanged in shape and grows somewhat stronger as external magnetic field is increased towards 5\,T for $\mathbf{H}||b$. No thermal hysteresis is found. Analogous measurements for $\mathbf{H}||a^\ast$, which suppresses the VC phase, show that ferroelectric polarization is still present in the VC phase and prefers the $b$-axis (see Supplemental Material).


\begin{figure}
\includegraphics[clip,width=0.785\linewidth]{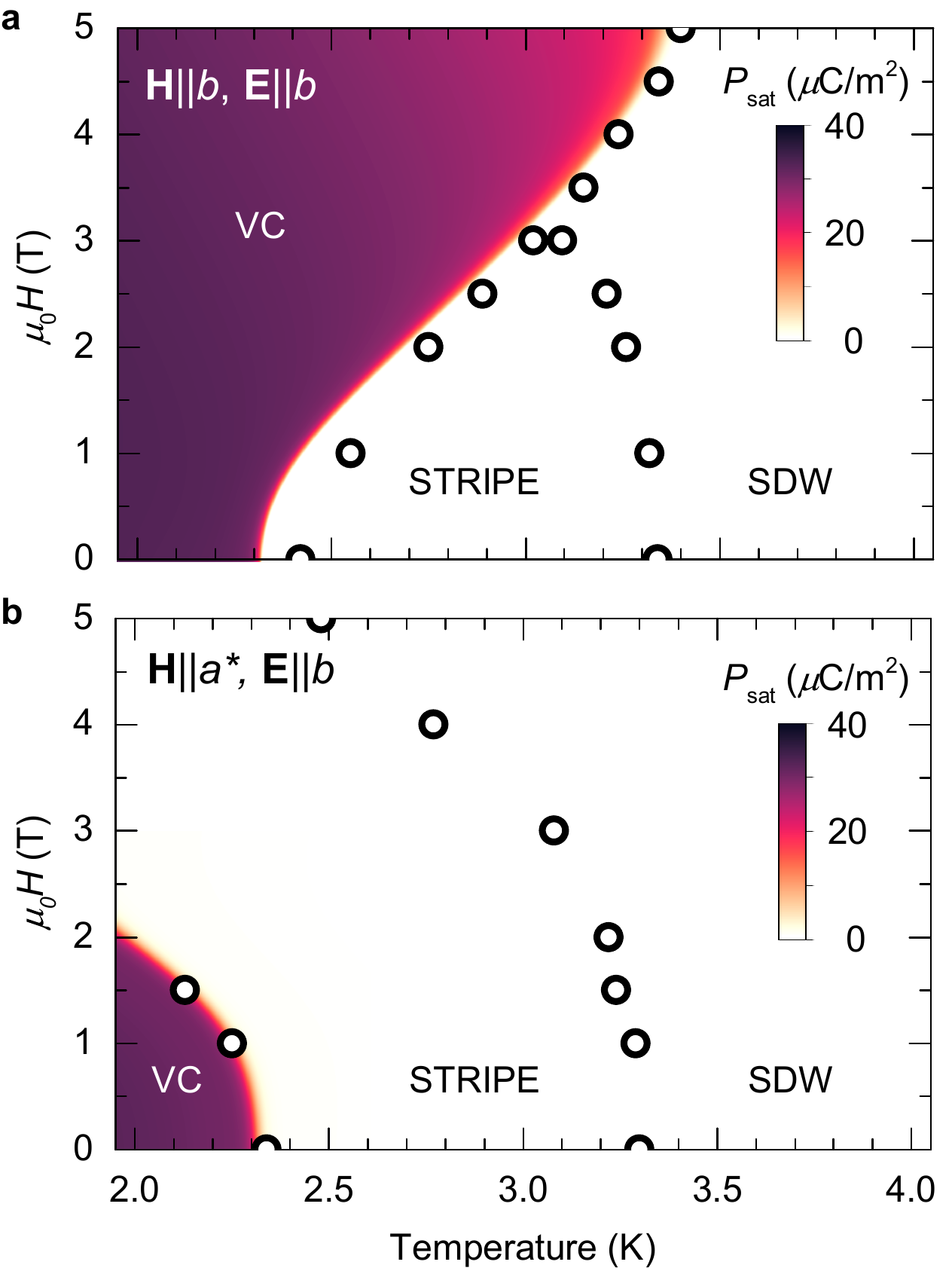}
\caption{Ferroelectric phase diagram of $\beta$-TeVO$_4$ for $\mathbf{H}||b$ (upper panel) and $\mathbf{H}||a\ast$ (lower panel) compared with critical temperatures of magnetic transitions. Shading represents interpolated measured saturation polarization $P_\textrm{sat}$ for $\mathbf{E}||b$, clearly demonstrating that the ferroelectric response appears only within the vector-chiral phase boundary. Circles denote critical temperatures determined by magnetic susceptibility from Ref.\ \onlinecite{savina2015}.}
\label{fig:phase_diag}
\end{figure}

We are now able to build the anisotropic ferroelectric phase diagram of $\beta$-TeVO$_4$. Fig.\ \ref{fig:phase_diag} compares the saturation polarization as a function of temperature and magnetic field with the magnetic phase diagram.\cite{savina2015} The onset of ferroelectricity closely follows $T_\mathrm{N3}$, i.e.\ the boundary of the VC phase, as $\mathbf{H}||b$ or $\mathbf{H}||a^\ast$ is applied. The ferroelectric response is thus exclusive to the VC order. Paraelectricity, if at all present, may exist in SDW and spin-stripe phases only in the vicinity of the VC boundary.

\begin{figure}
    \includegraphics[clip,width=0.825\linewidth]{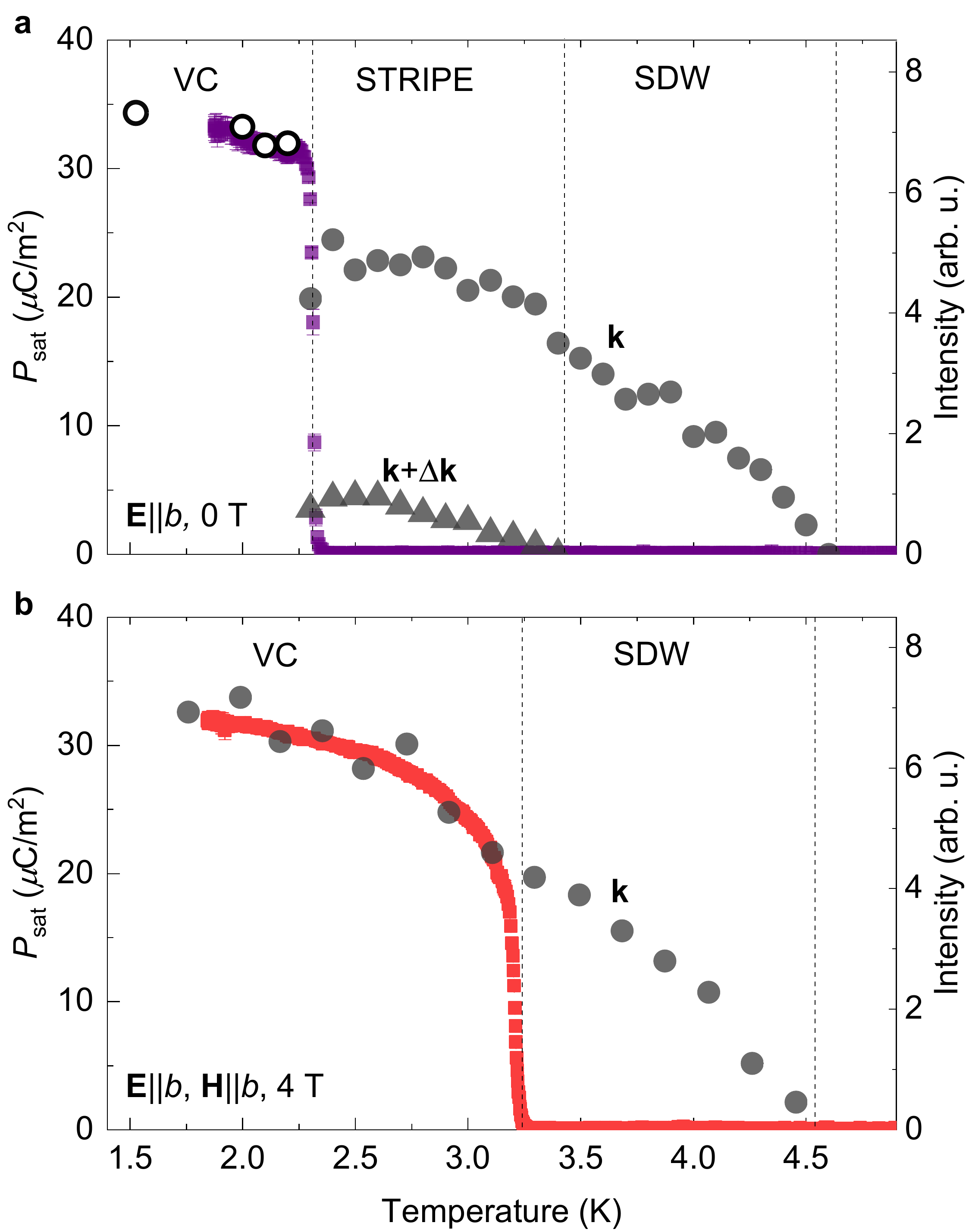}
    \caption{Saturation polarization of $\beta$-TeVO$_4$ compared to neutron diffraction intensities of  $\mathbf{k}=(-0.2, 0, 0.42)$ magnetic reflection and its satellite. (a) In $H=0$\,T the main magnetic reflection develops in the SDW phase at wave vector $\mathbf{k}$ (full circles) and in the stripe phase is accompanied with a satellite reflection at $\mathbf{k} + \Delta \mathbf{k}$ (triangles). Below $T_\mathrm{N3}$, $P_\textrm{sat}$ (purple squares) is proportional to diffraction intensity of the sole magnetic reflection at $\mathbf{k}$ (empty circles).  
    (b) For $\mathbf{H}||b$ at 4\,T only the main magnetic reflection (full circles) is observed and it is proportional to $P_\textrm{sat}$ (red squares) once the VC phase is established. Neutron diffraction intensities are from Ref.\ \onlinecite{pregelj2019magnetic}.}
    \label{fig:magnetic_parameters}
\end{figure}

The appearance of ferroelectricity, ie.\ the steep slope of $P_\mathrm{sat}$ {\it vs} $T$ in Fig.\ \ref{fig:diel}(c), becomes significantly less sharp when $\mathbf{H} || b$ is applied, depicted by the change of shading in Fig.\ \ref{fig:phase_diag}. In order to clarify the nature of this phase transition, it is useful to compare it with the formation of the long-range magnetic order. Neutron diffraction studies on $\beta$-TeVO$_4$ identified several magnetic reflections of interest.\cite{pregelj2015, pregelj2019elementary, pregelj2019magnetic} All of these reflections behave similarly with temperature and applied magnetic field, indicating that they probe the magnetic order parameter in a similar manner. To examine the  relation between the  magnetic order parameter and saturation polarization $P_\textrm{sat}$, we compare the temperature dependence of $P_\textrm{sat}$ with the intensity of the strongest magnetic reflection at $\mathbf{k}=(-0.2, 0, 0.42)$. As shown in Fig.\ \ref{fig:magnetic_parameters}(a), without applied magnetic field, the main reflection appears at $T_\mathrm{N1}$ and is associated with the formation of the SDW phase. Below $T_\mathrm{N2}$, a second orthogonally polarized SDW develops, which is seen as the appearance of the satellite reflection at $\mathbf{k} + \Delta \mathbf{k}$. At $T_\mathrm{N3}$, $\Delta \mathbf{k}$ locks to 0 as the satellite merges with the main reflection and the VC phase sets in. The ferroelectric polarization appears sharply at $T_\mathrm{N3}$ and is proportional to the total neutron scattering intensity at the wave vector $\mathbf{k}$. Apparently, even though both SDWs are present already above $T_\mathrm{N3}$, it is not until their locking, acting like a sharp trigger, that the electric dipoles form and immediately order ferroelectrically. The first-order character of this phase transition is supported by step-like changes of various magnetic properties.\cite{savina2015, weickert2016, pregelj2016} The resulting sharp jump in electric polarization found in $\beta$-TeVO$_4$ at $T_\mathrm{N3}$ is superficially similar to polarization flop induced by magnetic field\cite{Kimura2003} or to an incommensurate-commensurate magnetic transition.\cite{lawes2005, lawes2008} However, in those cases a thermal hysteresis is commonly present. For $\beta$-TeVO$_4$, no temperature hysteresis is observed neither in our data nor in previous literature. This implies the ferroelectric transition below the triple point is of the first order, with fluctuations of the order parameter likely being somewhat suppressed.

Furthermore, applying $\mathbf{H}||b$ above the triple point of 3.2\,T suppresses the spin stripe phase so that the VC phase borders the single-modulation SDW phase. As Fig.\ \ref{fig:magnetic_parameters}(b) shows for 4\,T, only one magnetic reflection is observed since the second SDW starts to gradually develop at $T_\mathrm{N3}$ with the same wave vector as the first SDW. This gradual onset of VC ordering is accompanied by a wide, gradual appearance of ferroelectric polarization which lacks a sharp jump visible at lower fields. Well below $T_\mathrm{N3}$, magnetic fluctuations subside with cooling and $P_\mathrm{sat}$ is proportional to the neutron scattering intensity. The  second-order-like character of the SDW-to-VC transition implies a continuous evolution of free energy minima corresponding to the VC order, with fluctuations close to $T_\mathrm{N3}$ being larger than below the triple point. This is, indeed, supported by the increased height of the dielectric peak above 3.2\,T [Fig.\ \ref{fig:diel}(c) inset].

A similar correlation between ferroelectricity and polarized neutron scattering intensities was observed within the spin-spiral order of orthorhombic TbMnO$_3$.\cite{Yamasaki2007} There, spontaneous electric polarization was found to be proportional to the difference of magnetic scattering intensities corresponding to two neutron polarizations. TbMnO$_3$ features polarization flop, which is absent in $\beta$-TeVO$_4$. Nevertheless, the ferroelectric transition above the triple point of $\beta$-TeVO$_4$ seems to be related to the general class of sinusoidal-to-helicoidal transitions in incommensurate ferroelectric magnets as proposed by Mostovoy.\cite{Mostovoy2006} We stress that $\beta$-TeVO$_4$ is a special case as it also prominently features the spin-stripe phase driven by the two superimposed incommensurate SDWs, which makes this system unique among multiferroics and magnetic ferroelectrics.

The above observations provide clear evidence that ferroelectricity and vector-chiral ordering of spins in $\beta$-TeVO$_4$ have a common origin. Also, we can conclude that electric measurements and neutron diffraction probe the same ordering entities in the VC phase with both magnetic and electric character. We now attempt to identify the mechanism responsible for establishing the ferroelectric polarization in the VC phase. The vector-chiral ordering in $\beta$-TeVO$_4$ comprises two elliptic spin spirals, each on a crystallographically independent zig-zag chain of vanadium.\cite{pregelj2016} Knowing that ferroelectric polarization prefers the $b$-axis, a likely mechanism for the polar order is the so-called inverse Dzyaloshinskii-Moriya (IDM) interaction where magnetic ordering may induce small shifts of ligand ions and create electric polarization.\cite{Katsura2005, Arima2011} It was previously used to predict ferroelectricity in, e.g., TbMnO$_3$\cite{Sergienko2006,Lu2015}, MnWO$_4$,\cite{Toledano2010} and MnSb$_2$S$_4$.\cite{De2018} In this case, the electric polarization $\mathbf{P}_\mathrm{IDM}$ for an incommensurate spin spiral structure is modeled as\cite{Katsura2005}
\begin{equation}
\mathbf{P}_\mathrm{IDM} \propto \sum_{\mathrm{pairs\ }(i, j)}\gamma \mathbf{e}_{ij} \times (\mathbf{S}_i \times \mathbf{S}_j),
\label{invDM}
\end{equation}
where $\gamma$ is a constant proportional to the spin-orbit coupling and superexchange interaction and  $\mathbf{e}_{ij}$ is the spatial vector connecting spins $\mathbf{S}_i$ and $\mathbf{S}_j$. We express each spin $\mathbf{S}_i$ in the $(a^\ast, b, c)$ coordinate system as $\mathbf{S}_i
= S_{ia^\ast}\hat{\mathbf{a}}^\ast + S_{ib}\hat{\mathbf{b}} + S_{ic}\hat{\mathbf{c}}$ and $\mathbf{e}_{ij} = e_{ij}\hat{c}$ (along the chains). Spins in the $bc$ cycloidal spiral of $\beta$-TeVO$_4$ can be written as $\mathbf{S}_i = S_{ib}\hat{\mathbf{b}} + S_{ic}\hat{\mathbf{c}}$ and their contribution to electric polarization is
\begin{equation}
\mathbf{P}_{\mathrm{IDM},bc} \propto \sum_{\mathrm{pairs\ }(i, j)} \gamma e_{ij} \left(S_{ib}S_{jc} - S_{ic}S_{jb}\right)\hat{\mathbf{b}}.
\label{eq:P_IDM}
\end{equation}
On the other hand, $S_c$ for the $ab$ spiral is negligible compared to $S_{a^\ast}, S_{b}$ since at low temperatures the monoclinic unit cell angle $\beta$ is $90.779^\circ$.\cite{pregelj2015} Hence the contribution $\mathbf{P}_{\mathrm{IDM},ab}$ of the $ab$ spiral is small. It follows that the dominant contribution to electric polarization under IDM model comes from the $bc$ spiral of $\beta$-TeVO$_4$ and the total electric polarization would point approximately along the $b$ axis, as also found experimentally (Fig.\ \ref{fig:diel}). The IDM model is also in agreement with our observation that $P_\mathrm{sat}$ is proportional to neutron scattering intensity, i.e.\ square of the magnetic order parameter or more likely to a product of two magnetic order parameters that contribute to the intensity of the same magnetic reflection, as found in some multiferroics.\cite{lawes2008}

In principle, an elliptically modulated spin spiral structure may allow also for the exchange striction mechanism that is driven by symmetric exchange interaction.\cite{greenwald1950,yahia2017} However, if this were responsible for ferroelectricity in $\beta$-TeVO$_4$ the electric polarization should appear already in the SDW phase which is obviously not the case. Another possible mechanism of spin-driven ferroelectricity could be the $p$-$d$ hybridization where, like in some delafossite,\cite{Terada2008,Fiebig2016} the spin-orbit coupling between the transition metal ion and the surrounding ligands might induce electric dipoles.\cite{arima2007_2,jia2007_2} The contribution of this mechanism is difficult to estimate because the low-temperature crystallographic structure of $\beta$-TeVO$_4$ is not precisely known. Finally, the center of spatial inversion may be lost due to lone pairs of Te$^{4+}$. In that case we would expect the polarization saturation\cite{Fiebig2016} as well as dielectric peak height\cite{Krohns2019} to be at least two orders of magnitude larger than observed. Since the observed polarization lies along the $b$-axis accompanied by the $bc$ spin cycloid, we conclude that the formation of electric dipoles in $\beta$-TeVO$_4$ is likely driven by the inverse Dzyaloshinskii-Moriya mechanism.

\begin{figure}
    \includegraphics[clip,width=0.95\linewidth]{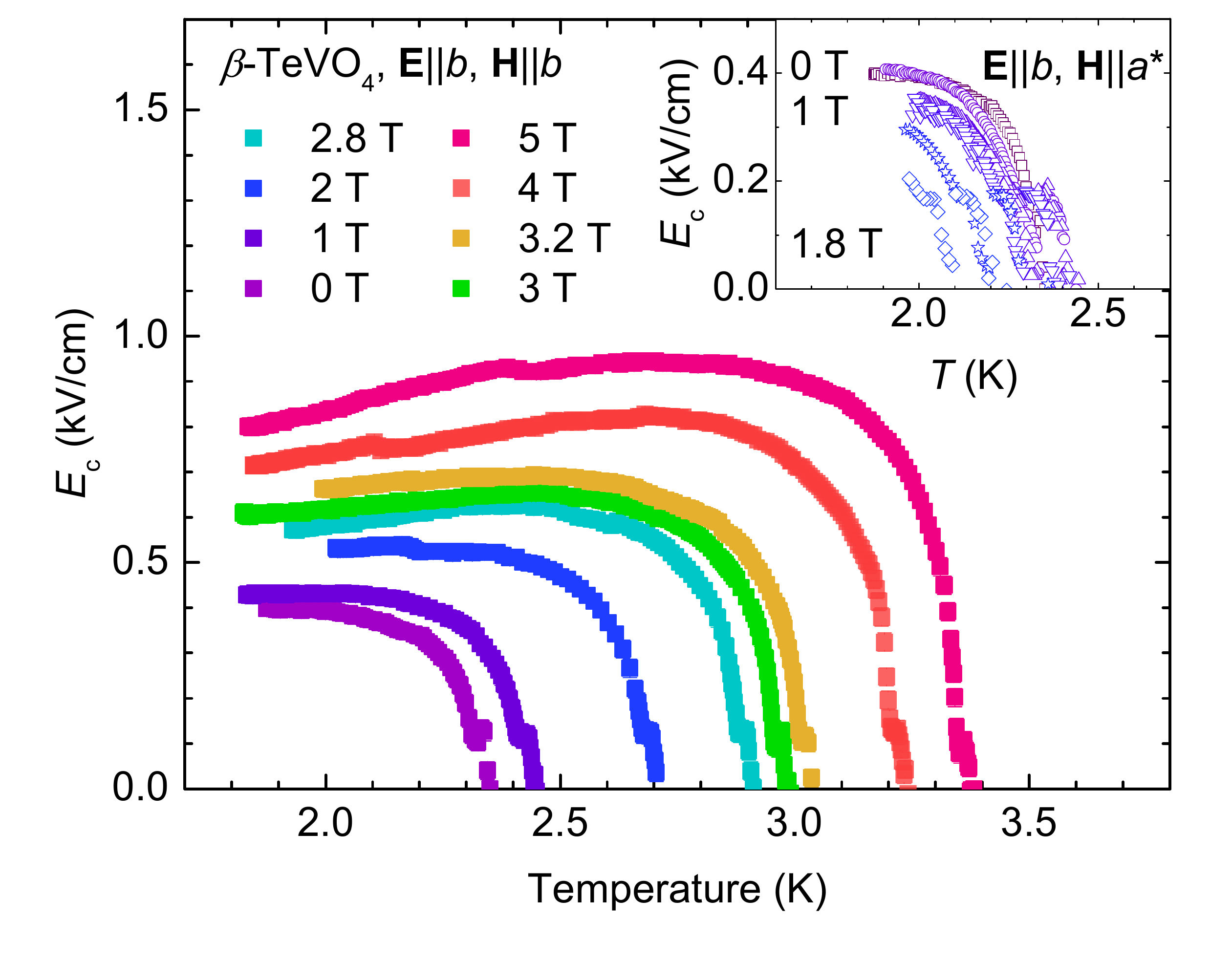}
    \caption{Coercive field of ferroelectric hysteresis $E_\textrm{c}$ as a function of temperature and applied magnetic field $\mathbf{H}||b$ (main panel), $\mathbf{H}||a^\ast$ (inset). The increase of $E_\textrm{c}$ at high magnitudes of $\mathbf{H}||b$ indicates that magnetoelectric coupling terms of higher order are present, see text.}
    \label{fig:coercive}
\end{figure}

The applied magnetic field does not significantly influence the magnitude or anisotropy of the low-temperature polarization which rules out strong linear magnetoelectric coupling in $\beta$-TeVO$_4$. Assuming that the IDM mechanism is dominant, this means that both independent spirals of vanadium spins remain within their respective planes as the magnetic field increases. Indeed, our previous findings confirm that for $H \leq 5$\,T  there is no macroscopic field-induced spin reorientation in the VC phase.\cite{herak2020} In sharp contrast, a clear electric polarization flop is observed in MnWO$_4$ and TbMnO$_3$ due to the field-induced reorientation of spiral spins.\cite{taniguchi2006,Kimura2003}

However, a certain form of nonlinear magnetoelectric coupling is nevertheless visible. Fig.\ \ref{fig:coercive} shows that as $\mathbf{H}||b$ is applied there is a remarkable increase of ferroelectric coercive field $E_\textrm{c}$, the field required to force bulk ferroelectric polarization to zero, determined as half width of the $P_\mathrm{FE}$--$E$ loops. As the VC phase is promoted, the ferroelectric hysteresis hardens significantly, e.g., at 2\,K, $E_\textrm{c}$ doubles when $\mathbf{H}||b$ is increased from 0\,T to 5\,T. In contrast, $\mathbf{H}||a^\ast$ suppresses the VC phase but otherwise does little to change $E_\textrm{c}$ except from a temperature shift (Fig.\ \ref{fig:coercive} inset). The magnetic dependence of $E_\textrm{c}$ in $\beta$-TeVO$_4$ is surprising since below $T_\mathrm{N3}$ the magnetic order deep in the VC phase appears rigid and fully established regardless of $\mathbf{H}||b$.\cite{herak2020} We also note that the applied electric field does not change the macroscopic magnetic susceptibility anisotropy within the VC order (see Supplemental Material). Nevertheless, a dependence of $E_\textrm{c}$ on $H$ is in principle allowed for magnetoelectric ``weak ferroelectrics'' and is a sign of higher-order magnetoelectric coupling terms of $EHH$ type,\cite{Schmid1973,Schmid2008} which in fact is reflected in the presence of IDM magnetoelectric coupling.\cite{Eerenstein2006} We may tentatively relate the enhancement of $E_\mathrm{c}$ in $\beta$-TeVO$_4$, as VC phase is promoted, to MnWO$_4$ where the applied magnetic field destabilizes the chiral phase and the ferroelectric coercive field $E_\textrm{c}$ decreases.\cite{Kundys2008} However, a word of caution is needed as MnWO$_4$ concomitantly undergoes a polarization flop.\cite{taniguchi2006} This is fundamentally different to $\beta$-TeVO$_4$ where the spin structure in the VC phase shows no signs of spin flops or reorientation of ferroelectric polarization.

Our results suggest a different physical picture in which the population of chiral domains, coinciding with ferroelectric domains, may be controlled via electric field. Even though the applied magnetic field does not influence the antiferromagnetic domain structure, in the context of IDM mechanism reversing the polarization of a ferroelectric domain requires reversing its spin chirality. Once the VC order is established, we directly observe that the applied $\mathbf{H}||b$ increases the energy barrier between the two types of ferroelectric-chiral domains in $\beta$-TeVO$_4$, i.e., the domain walls become less mobile as the VC phase is promoted. In a ferroic system, a domain wall with lower mobility is generally found to be thinner.\cite{choudhury2008} We thus propose that in $\beta$-TeVO$_4$, when magnetic field $\mathbf{H}||b$ promotes the VC phase, intermediate spin configurations inside the thin domain wall become unfavorable and polarization reversal likely happens only via tunneling between the two chiral states, effectively increasing $E_\textrm{c}$.

The microscopic mechanism of external control of $E_\mathrm{c}$ by the applied magnetic field in single-crystal spin-driven multiferroics has not been researched in detail. This is in contrast to composites, inhomogeneous polycrystalline, or thin film systems with freely rotating or flopping dipoles such as e.g., the iron-polyvinylidenefluoride-iron heterostructure\cite{carvell2013} or the DyMnO$_3$\cite{Lu2013} where magnetoelectric coupling is mediated indirectly by lattice strain, magneto-, or electrostriction. Indeed, magnetostriction was found in $\beta$-TeVO$_4$, the largest effect being at the transition to VC magnetic order.\cite{weickert2016} Currently it is unclear to which extent does magnetostriction within the domain walls play a role in establishing and modulating ferroelectric properties of $\beta$-TeVO$_4$. It therefore seems worthwhile to examine other external parameters such as pressure or strain that may affect ferroelectricity in bulk weak multiferroics, as well as to explore the magnetoelectric effect in $\beta$-TeVO$_4$ at higher magnetic fields, up to 20\,T, where the vector-chiral ordering becomes suppressed.\cite{pregelj2015,pregelj2019magnetic,weickert2016}

In conclusion, we have presented a detailed examination of ac dielectric response and static electric polarization of the antiferromagnet $\beta$-TeVO$_4$ with magnetic and electric fields applied along various crystallographic axes. Within the vector-chiral phase the ferroelectric polarization is discovered which confirms that spatial inversion symmetry is broken by the spiral arrangement of spins. The anisotropy of the spontaneous electric polarization is consistent with the inverse Dzyaloshinskii-Moriya mechanism. Saturation polarization follows the behavior of neutron scattering intensities corresponding to the magnetic order parameters.  This observation allows us to characterize the spin-stripe-to-VC and the SDW-to-VC phase transitions as first-order- and second-order-like, respectively.  Finally, we discuss an unexpectedly strong influence of the applied magnetic field on the width of the ferroelectric hysteresis, i.e., magnetic-field-induced pinning of the multiferroic domains. Control of ferroelectric coercivity via manipulation of spin-chiral energy landscape may have far-reaching implications for stability and density of stored ferroelectric-chiral information. Our findings underline the potential of spiral-like spin orderings in creating and controlling robust multiferroic phases.

\begin{acknowledgments}
We thank Ivan Balog for elucidating discussions. This work was supported by the Croatian Science Foundation, Grants No.\ IP-2018-01-2730 and IP-2013-11-1011. M.H.\ acknowledges support by the Croatian Science Foundation Grant No.\ UIP-2014-09-9775. M.P., A.Z., and D.A.\ acknowledge financial support of the Slovenian Research Agency through Grants No.\ P1-0125, J1-9145, J1-2461, and J2-2513.
\end{acknowledgments}

\bibliographystyle{apsrev4-1}
\bibliography{REF}

\end{document}